# In situ high-cycle fatigue reveals the importance of grain boundary structure in nanocrystalline Cu-Zr

Jennifer D. Schuler [a,b], Christopher M. Barr [b], Nathan M. Heckman [b], Guild Copeland [b], Brad L. Boyce [b], Khalid Hattar [b], Timothy J. Rupert [a,*]

[a] Department of Materials Science and Engineering, University of California, Irvine, CA 92697, USA

[b] Material, Physical, and Chemical Sciences, Sandia National Laboratories, Albuquerque, New Mexico 87185, USA

[*] Corresponding Author: trupert@uci.edu

**Abstract**

Nanocrystalline metals typically have high fatigue strengths, but low resistance to crack propagation. Amorphous intergranular films are disordered grain boundary complexions that have been shown to delay crack nucleation and slow crack propagation during monotonic loading by diffusing grain boundary strain concentrations, suggesting they may also be beneficial for fatigue properties. To probe this hypothesis, in situ transmission electron microscopy fatigue cycling is performed on Cu-1 at.% Zr thin films thermally treated to have either only ordered grain boundaries or to contain amorphous intergranular films. The sample with only ordered grain boundaries experienced grain coarsening at crack initiation followed by unsteady crack propagation and extensive nanocracking, whereas the sample containing amorphous intergranular films had no grain coarsening at crack initiation followed by steady crack propagation and distributed plastic activity. Microstructural design for control of these behaviors through simple thermal treatments can allow for the improvement of nanocrystalline metal fatigue toughness.

## Introduction

While nanocrystalline metals, defined as having an average grain size less than 100 nm, have excellent structural properties such as high strength [1], hardness [2], and wear resistance [3], these properties are challenged by the most widespread mechanical failure: fatigue [4]. Nanocrystalline metals can usually achieve longer overall fatigue lifetimes compared to coarse-grained counterparts [5], but their weakness is limited resistance to crack growth and hence rapid failure following crack nucleation [6]. Fatigue lifetime in the high-cycle, low-amplitude regime can be considered in two stages as (1) crack initiation, followed by (2) crack propagation until sudden failure. Crack initiation in nanocrystalline metals has been shown to be preceded by abnormal grain growth and slip protrusions [7-9], whereas coarse-grained metal crack initiation relies primarily on persistent slip band activity [10]. Once initiated, cracks propagate through combinations of plasticity and interior crack formation that are dependent on the loading conditions and grain size [11, 12], driven by mechanisms such as dislocation nucleation and motion [13], deformation twinning [14], grain boundary migration [15], grain boundary sliding [16, 17], cooperative grain rotation [18], and cavitation [13]. Crack propagation in coarse-grained metals is resisted by tortuosity, plasticity, and roughness-induced crack closure, but these mechanisms all become suppressed with decreasing grain size [11].

Complexions are defined as thermodynamically-stable grain boundary features that can assume a range of ordered or disordered structures [19], where the disordered version with an equilibrium thickness would be called an amorphous intergranular film (AIF). Nanocrystalline grain sizes can potentially offer new opportunities by leveraging their associated high grain boundary volume fraction [20] through complexions. Nanocrystalline Cu-Zr alloys with AIFs have both increased strength and ductility compared to the same alloy with only conventional, ordered grain boundaries [21]. AIFs increase ductility and damage tolerance by diffusing the local strain concentrations at the grain boundary caused by dislocation absorption, which results in slower crack nucleation and growth [21, 22]. To date, this behavior has only been studied under monotonic loading scenarios, but it is hypothesized that a similar mechanism may also allow for improved fatigue behavior.

In this study, in situ transmission electron microscopy (TEM) fatigue was performed on nanocrystalline Cu-1 at.% Zr thin films processed to have either only ordered grain boundaries ("ordered grain boundary sample") or to contain AIFs scattered throughout the grain boundary network ("AIF-containing sample"). Microstructural analysis during crack initiation and propagation reveals grain growth at crack initiation with unstable crack growth and extensive nanocracking in the ordered grain boundary sample, whereas the AIF-containing sample had no grain growth at crack initiation with stable crack growth and distributed plasticity. Both samples were extrinsically toughened through grain bridging, while the more ductile AIF sample was also toughened intrinsically through crack tip plasticity mechanisms. By tuning grain boundary structure and composition, traditional extrinsic and intrinsic fatigue toughening mechanisms can be applied to resist crack propagation in nanocrystalline metals.

## Materials and Methods

Cu-1 at.% Zr thin films were produced with magnetron co-sputtering using an Ar plasma with a Kurt J. Lesker Lab 18 modular thin film deposition system. The films were deposited at room temperature using 400 W for Cu and 46 W for Zr at 12 mT onto polished salt substrates. The films were then floated onto Protochips, Inc. Fusion heating chips through dissolution of the salt substrate in a solution of water and isopropyl alcohol. Additional details describing the film deposition and sample preparation are provided in Supplementary Note 1. The films were annealed under vacuum using an Aduro double tilt heating holder in an FEI Tecnai G(2) F30 S-Twin 300 kV TEM. The samples were subjected to different heat treatments on separate heating chips in order to create the ordered grain boundary or AIF-containing samples. First, both samples were heated from 25 to 850 °C over the course of 1 h (ramp rate of 0.23 °C/s) and then held at 850 °C for 1 h. AIFs have been previously observed in ball milled Cu-Zr alloys that were annealed at this temperature [23]. After the 1 h anneal, one sample was slowly cooled over the course of 600 s to 25 °C (cool-down rate of 1.4 °C/s) to create the ordered grain boundary sample. The AIF structure is only stable at high temperatures where grain boundary pre-melting occurs, so the slow cooling treatment allows these features to crystallize and form the typical ordered grain boundary structure. Alternatively, another sample was quickly cooled in <1 s by turning off the applied electrical current, which rapidly removes the heating input. Since the remainder of the heating chip is much larger than the sample itself, the specimen rapidly cools and "freezes" in any boundary structures that were stable at high temperatures (AIFs for these alloys) [24]. This rapid cooling step is analogous to quenching of powder samples that was used previously for ball milled Cu-Zr [23]. Similar rapid quenching to freeze in an amorphous structure has also been used to create monatomic metallic glasses [25]. Bright field TEM, high resolution TEM, selected area electron diffraction (SAED), and scanning TEM energy-dispersive X-ray spectroscopy (STEM-EDS) were performed with a JEOL 2800 and JEOL 2100 operated at 200 kV. A film thickness of 51 ± 6 nm was measured for the as-deposited sample through electron energy loss spectroscopy (EELS) at 300 kV in a JEOL Grand ARM300CF, with additional EELS measurement details in Supplementary Note 2. Average grain sizes were calculated from bright field TEM images by measuring the equivalent circular diameters of at least 100 grains from each sample.

The in situ TEM fatigue methodology was modeled after a prior study by Bufford et al. [26]. Electron transparent Single Edge Notched Tension (SENT) [27] specimens were prepared from the annealed thin films using a focused ion beam (FIB) lift-out technique with an FEI Nova 600 Nanolab FIB scanning electron microscope (SEM), where efforts were made to minimize inadvertent Pt deposition and FIB damage over the gauge area, both of which can impact mechanical behavior [28]. These specimens were then placed on push-to-pull (PTP) devices from the Bruker Corporation for testing. Additional details regarding mechanical testing sample preparation are provided in Supplementary Note 1. The PTP devices were actuated with a Bruker PI 95 TEM PicoIndenter holder in a JEOL 2100 operated at 200 kV at Sandia National Laboratories [29]. Tension-tension open loop fatigue was performed at 100-200 Hz with a staircase loading regimen [30] consisting of progressively increasing peak loads and amplitudes, with the full loading conditions provided in Supplementary Note 3. The loading conditions between the two samples were identical until final fracture of the AIF-containing sample, with the ordered grain boundary sample receiving continued progressive loading until fracture. While the final loading conditions differ between the two samples, our focus was on ensuring that the fatigue crack progressed so that we could understand the relevant deformation mechanisms for each material. The fatigue tests were performed with continuous video recording using a 680 nm × 680 nm field of view and an image capture rate of 15 frames per second in bright field TEM mode that was optimized for diffraction contrast. Note that during fatigue cycling, TEM images were blurred since the loading frequency was faster than the frame rate, with approximately 14 cycles captured per frame. In order to better capture the evolution of each sample, cycling was interrupted by periods of constant load where cycling was stopped and the samples partially unloaded to capture clear images. The full fatigue test videos are provided in the Supplementary Information, accelerated to 10 times the normal speed for ease of viewing. All bright field TEM images of the fatigue specimen captured from the video have been rotated so that the loading direction is vertical, and crack propagation is from left to right.

**Results and Discussion**

Bright field TEM images of the ordered grain boundary sample and AIF-containing sample are shown in Figs. 1(a) and (b), respectively, where the white arrow in Fig. 1(a) indicates the loading direction. Fig. 1(c) shows an SEM image of the sample preparation on the PTP device using Pt deposition. The brighter contrast over the gauge region indicates that the deposited thickness of the thin film is sufficiently thin as to be electron transparent. Experimental factors that are hard to control such as precise sample geometries, notch shape, specimen warp, internal film stress, changing grain imaging conditions, and imperfect Pt attachment at the gripper region limit the potential discussion of exact fatigue properties. Even within a given sample, specimen thickness variations preclude direct interpretation of quantitative data. However, differences in microstructural mechanisms and behavior during fatigue are investigated and reveal insight as to how AIFs can contribute to fatigue crack growth resistance. Bright field TEM images of the ordered grain boundary (Fig. 1(d)) and AIF-containing (Fig. 1(e)) samples are shown at the regions of interest immediately adjacent to the notch where crack initiation and propagation occurred. Locally thin regions are present in both films, which may be due to the sputter deposition process [31], grain boundary grooving, or preexisting inhomogeneities. The average grain size of the ordered grain boundary sample is 43 ± 11 nm, while the average grain size of the AIF-containing sample is 46 ± 14 nm, meaning that only one or possibly two grains are through the film thickness. A few abnormally large grains at the notch, visible in Fig. 1, are present prior to fatigue and are likely due to FIB-induced grain growth from notch creation [32]. The SAED pattern insets in Figs. 1(d) and (e) only have Cu face centered cubic rings, overlaid in blue, meaning no detectable second phase such as hexagonal close packed Zr or an intermetallic was observed. The film composition of 0.9 ± 0.3 at. % Zr was measured by averaging values across STEM-EDS line scans. Clear evidence of dopant segregation to the grain boundaries was not detected unlike previous Cu-Zr experimental studies [23, 33]. However, the solid solubility of 0.12 at.% Zr in Cu coupled with compositions near the EDS resolution lower limits may make dopant segregation in this study more subtle and difficult to measure than in previous experimental studies that used higher Zr doping percentages. The high resolution TEM image in Fig. 1(f) shows an AIF from the quenched film that was identified through Fresnel fringe imaging [34] in edge-on conditions. AIF formation is dependent on the grain boundary energy, local chemistry, and quenching rate, so it does not occur at every grain boundary [19, 21]. Instead, AIF-containing samples have a distribution of complexion types, including both ordered and disordered grain boundaries.

Analysis of crack initiation, the first stage of fatigue damage, is presented first. Crack initiation is defined as when a crack ingress has been observed beyond the FIB-created notch. In both the ordered and AIF-containing grain boundaries, the nanoscale fatigue cracks first initiated about one grain diameter (~20 nm) ahead of the notch, presumably due to the higher stresses ahead of the notch and/or weaker microstructural features to enable crack initiation. Grain growth preceded crack initiation in the ordered grain boundary sample, reminiscent of prior nanocrystalline metal fatigue studies [7-9], but grain growth was absent during crack initiation in the AIF-containing sample. AIFs strongly resist grain growth, so much that even when nanocrystalline ball milled Cu-Zr containing AIFs was held at 98% of its solidus temperature for a week, it remained nanocrystalline [23]. In fact, Schuler et al. [35] even observed a new regime of high temperature nanocrystalline stability due to AIF formation in Ni-W alloys. The ability of AIFs to both stabilize the grain size and diffuse local grain boundary strain concentrations may contribute to the absence of grain coarsening in the AIF-containing sample at fatigue crack nucleation. Figs. 2(a)-(f) show crack initiation at the notch marked by the dashed line in the ordered grain boundary sample with progressive cycling in the loading direction indicated by the white arrow in Fig. 2(a). The number of elapsed cycles is indicated in white at the bottom right of each frame. The grain denoted by the black arrows in Figs. 2(a)-(f) more than triples in size from 12 nm to 41 nm. Although the grain growth is less dramatic than that observed in other nanocrystalline metal fatigue studies [7, 8, 26], this may be due to the maximum grain size being limited by the thin film thickness [36]. In the ordered grain boundary sample, black contrast features were observed to migrate during cyclic loading, particularly across the grains marked by the black and blue arrows in Figs. 2(e) and (f), and can be more fully seen in Supplementary Video 1. While the defect character was not determined due to the unknown beam condition and diffraction contrast, the contrast is consistent with dislocation-based plasticity [14, 31]. The red arrows in Fig. 2(f) mark competing crack initiation sites where intergranular cracks have formed. The grain adjacent to the lower interior crack in Fig. 2(f) eventually yields, allowing the intergranular crack to connect to the notch and commence crack propagation. In contrast, the AIF-containing sample shown in Figs. 2(g)-(l) had distributed plastic activity evidenced by discrete microstructural contrast changes surrounding the notch region until an internal crack marked by the red arrow in Figs. 2(k) and (l) formed at a microstructural feature. The adjacent grain indicated by the blue arrow in Figs. 2(k) and (l) eventually breaks away, connecting the internal crack to the notch to allow crack propagation. Such interior crack formation and nanocracking has been previously observed as a primary crack propagation mechanism in nanocrystalline metals [11, 14, 37], where void formation at grain boundaries and triple junctions ahead of the main crack contribute to intergranular fracture [38].

Crack propagation, the second stage of fatigue lifetime, is analyzed next. Grain bridging is defined as when a grain spans the opposing fracture surfaces in the crack wake, dissipating energy that would have extended the crack tip [39, 40]. A network of nanocracks and grain bridging events cause unstable crack propagation in the ordered grain boundary sample, whereas the AIF-containing sample had steady crack propagation punctuated by a series of nanocracks that ultimately connect to cause failure. Crack progressions captured from video frames are shown in Figs. 3(a)-(f) for the ordered grain boundary sample and Figs. 3(g)-(l) for the AIF-containing sample, with the number of elapsed cycles in the loading direction indicated in the bottom right corner of each image. Some small amount of relative motion of microstructural features in both samples visible in Fig. 3 may be due in part to global elongation caused by creep from the extended time under tension necessary to reach high-cycle fatigue. Similar microstructural shifts were also observed by Bufford et al. [26] during in situ TEM Cu high-cycle fatigue. The last frames before failure for the ordered grain boundary sample (Fig. 3(f)) and AIF-containing sample (Fig. 3(l)) show the ordered grain boundary sample failing during constant loading between cycling steps and the AIF-containing sample failing during active fatigue cycling, which causes the image to be blurry.

Unlike early fatigue studies in the TEM [41], the recently developed nanoindentation-based capability permits quantitative measurement of the driving force and crack advance throughout the fatigue test. From the measured forces and estimates for sample dimensions, it is possible to calculate approximate values for the plane stress linear elastic stress intensity range during fatigue loading. After the first 100 nm of propagation, the crack is expected to have escaped the influence of the ~90 nm notch radius and a rudimentary estimate of the stress intensity factor for the clamped SENT geometry is possible [42]. The crack tip is defined as the furthest ingress of the crack. Note that the crack tip could be a nanocrack with bridging grains in its wake [43]. For the AIF-containing specimen at a total crack length of 1.1 μm (for notch plus crack length *a*, which we note is a definition used only for this exercise, and specimen width *W*, $a/W$=0.33), the 10 μN applied force amplitude corresponds to a stress intensity factor range of $\Delta K$ ~ 0.4 MPa√m, well below the macroscopic fatigue threshold for Cu [44]. Direct in situ visual measurements of total crack length as a function of cycles shown in Figs. 4(a) and (b) for the ordered grain boundary and AIF-containing samples respectively enabled a determination of the crack growth rate, with additional details provided in Supplementary Note 3. For these conditions, we measure a crack growth rate of $2\times10^{-12}$ m/cycle for the AIF-containing sample. Given the yield strength of approximately 1 GPa [21], the corresponding plane stress monotonic plastic zone size is estimated to be in the vicinity of 35-65 nm, confirming small-scale yielding and a valid estimate of the plane stress $K_I$. This extremely low crack growth rate is comparable to the value reported previously for in situ measurements on pure Cu [26] and is a growth rate that is difficult to measure by other macroscopic test techniques. The low crack growth rate corresponds to a single lattice parameter of average crack advance every ~200 cycles. Consistent with direct observation, the crack grows intermittently, arresting and restarting at the atomic scale, in spite of the apparent monotonic growth shown in Fig. 4. The early crack growth rate for the ordered boundary specimen was even lower at ~$5\times10^{-13}$ m/cycle, in spite of a higher driving force $\Delta K$ ~ 1.2 MPa√m at a crack length of 1.1 μm. Finally, the crack growth accelerates in the ordered grain boundary sample, consistent with an increasing driving force for propagation as the crack propagated and the loading conditions increased. In contrast, the AIF-containing sample showed an unexpected constant crack growth rate in spite of the increasing driving force associated with both a growing crack and increasing load amplitudes. The growth rate behavior during in situ fatigue loading of metals warrants further investigation, since the observations reported here are substantially different from those reported for bulk ultrafine grained Cu [44].

Plastic activity was identified dynamically through observation in the video in frame by frame analysis, where moving dislocations and grain boundaries differentiated from bend contours through their discrete and asynchronous motion [31]. Instances of plastic activity, which may include dislocation nucleation, dislocation movement, or grain boundary migration, as a function of distance from the advancing crack tip are shown as heat maps in Fig. 5(a) for the ordered grain boundary sample and in Fig. 5(b) for the AIF-containing sample. In this figure, activity closest to the crack tip is black and the farthest is white. The background bright field TEM images show the last clear frame before sudden failure for reference. The load amplitude for all except the first 160,000 cycles in both samples corresponds to ~50 nm of displacement, causing a 1 to 2 grain ambiguity in the recorded plastic activity locations, corresponding to a ~40-80 nm potential error in the position measurements. The crack tip coordinates were then found during constant loading conditions between cycling sets and used to calculate the linear distance from the plastic event to the crack tip position at the time of occurrence. Cycling steps 44, 45 and 53 for the ordered grain boundary and 23 for AIF-containing samples, as described in Supplementary Note 3, were not analyzed for plastic behavior due to excessive drift. Plastic events were then separated as being in front of or behind the crack tip, such as at a grain bridge. 75% of the total plastic events in the ordered grain boundary sample were in front of the crack tip compared to 98% of AIF-containing sample plastic events, signifying enhanced plasticity preceding the crack tip when AIFs are present. The heat maps show that plastic activity in front of the crack tip is concentrated along the path of crack advancement for the ordered grain boundary sample, with localized clusters at the point of initiation and where a grain bridge eventually occurs. In contrast, plastic activity in the AIF-containing sample is more evenly distributed and far in front of the crack tip. Fig. 5(c) presents the position data from Figs. 5(a) and (b) as cumulative distributions. One can pick an arbitrary distance to understand the difference in the two distributions. For example, 50% of plastic activity in the ordered grain boundary sample occurred within ~130 nm from the crack tip, whereas 50% of plastic activity in the AIF-containing sample occurred within ~300 nm from the crack tip. Also, plastic activity only extended to ~600 nm from the crack tip in the ordered grain boundary sample, but reached up to 800 nm away in the AIF-containing sample.

For the ordered grain boundary sample, very few plastic events were recorded at nanocracks outside of the main crack, suggesting these features likely formed through sub-critical cleavage. Similar events were observed previously for sputtered thin films [31]. The extensive nanocrack network observed in the ordered grain boundary sample, but not the AIF-containing sample may be accounted for by the larger distribution of plastic activity in the AIF-containing sample. AIFs diffuse grain boundary strain, giving grain boundaries with these features a higher damage tolerance than a comparable ordered grain boundary [22]. The distributed plastic activity observed in the AIF-containing sample is likely due to AIFs mitigating boundary damage and allowing observable plastic activity to manifest, whereas the ordered grain boundary sample succumbed to nanocracking before having observable plastic activity.

The unstable crack propagation in the ordered grain boundary sample propagated through the formation of nanocracks and grain bridges. The evolution of one such grain bridge with progressive cycling is shown in Fig. 6(a) for the ordered grain boundary sample. The red arrow marks a grain that sustained considerably localized plasticity and has grown across the grain bridge, serving as the point of eventual detachment. While the AIF-containing sample also had grain bridging, it was accompanied by significant crack tip plasticity with steady crack propagation. Bright field TEM images of the AIF-containing sample in Fig. 6(b) shows a distinct “V” shape preceding the crack tip, with one example indicated by the red arrows, which may be a blunted crack or plastic hinge caused by local thinning and deformation from strain fields ahead of the crack tip [45-47], or movement of material through dislocation emission and absorption between nearby grain boundaries and the advancing crack tip [12]. The plastic deformation zone was also confined to the grains located immediately in front of the crack tip, indicating that possible slip transfer was limited by factors such as grain boundary character, slip system orientation, and angle of crack deflection [31, 48-51].

Next, analysis of the fracture surfaces post failure is presented. Tortuosity can be defined as the ratio between the total crack length and the distance between the crack starting and ending points, excluding the notch [52]. Propagation refers to the stage of crack growth after initiation until the critical length that causes sudden, uncontrolled failure is reached. Failure refers to the portion of the fracture surface formed after the propagation stage at sudden failure. Saw-toothing is defined as individual grains that undergo severe local plastic deformation and become ligaments until finally necking down to a point [14]. Bright field TEM images of the fracture surfaces are shown in Fig. 7 for the ordered grain boundary and AIF-containing samples. The dashed red lines mark the start of the propagation stage where crack nucleation occurs on the left and the crack propagates to the right. The solid blue lines mark the end of the propagation stage and the commencement of sudden failure. The fracture surfaces from the failure stages are shown in greater detail for the ordered grain boundary and AIF-containing samples in Figs. 7(c) and (e), respectively, with the outlines of the fracture surfaces shown in Figs. 7(d) and (f) to help guide interpretation. The crack deflection segment lengths, deflection angles, and tortuosity measurements from the fracture surfaces for each stage are presented in Table I. The larger average deflection angle in the failure versus propagation stage for both samples can be attributed to saw-toothing that was only observed in the failure stage, as shown in Figs. 7(c) and (e). The mean deflection lengths are on the same scale as the grain size, with good alignment between mating surfaces excluding saw-toothed regions [53]. The tortuosity is comparable between all stages, except the propagation stage in the AIF-containing sample that is significantly smoother with almost no measurable tortuosity. Fracture modes due to cyclic loading differ from monotonic loading conditions when crack tip plasticity is appreciable. Since plasticity is largely absent from brittle materials, the fracture surface morphologies subject to cyclic or monotonic loads will be similar in classically brittle materials such as ceramics [4]. The difference in tortuosity between the propagation and failure fracture surfaces observed in the AIF-containing sample are therefore another sign of enhanced ductility.

Toughening mechanisms can be categorized based on where they occur in relation to the crack tip and the shielding mechanisms. Extrinsic toughening operates behind the crack tip and lowers the effective force felt by the crack tip. Intrinsic toughening operates in front of the crack tip primarily through plasticity and normally operates in more ductile materials [4]. Plasticity-induced toughening was more extensive in the case of the ordered grain boundary sample – a result that was indirectly confirmed by the longer stable (subcritical) crack length prior to catastrophic failure in the ordered grain boundary sample compared to the AIF-containing sample. The effects of plasticity are also apparent in the propagating crack tip shape, as the ordered grain boundary sample showed a more open, blunted crack whereas the AIF-containing crack had a narrower crack profile, and correspondingly lower crack-tip opening displacement (Fig. 3). This enhanced plasticity-induced toughening is due, in part, to the 12% lower yield strength of the ordered grain boundary sample (938 MPa for ordered grain boundaries compared to 1068 MPa for AIF-containing) [21]. Grain bridging, an extrinsic mechanism, was present in both samples and can also contribute to improved toughness. Factors such as dopant segregation, grain boundary character, disordering, and energy state have been found to impact the damage tolerance of a grain boundary [54-59]. For example, minimizing the number of low-angle grain boundaries and enhancing twinning improves fracture toughness [54, 60, 61]. Incorporation of these techniques with extrinsic mechanisms to enhance grain bridging offers pathways to resist crack propagation in nanocrystalline metals. Alloys with appropriate doping and annealing conditions that permit AIF formation should utilize the enhanced ductility observed in this study to intrinsically toughen nanocrystalline alloys and avoid catastrophic, sudden fracture. Nanocrystalline alloys containing AIFs are also stronger than the same alloy with only ordered grain boundaries [21, 23, 62], offering a unique combination of ductile crack tip shielding with strengths even higher than nanocrystalline metal expectations.

**Summary and Conclusions**

Cu-1 at.% Zr thin films were thermally processed to have either ordered grain boundaries or contain AIFs and then subjected to in situ TEM fatigue. A number of observations are made, with our main findings categorized by fatigue lifetime stage.

1) *Crack initiation*: The ordered grain boundary sample experienced grain growth and dislocation activity at the crack initiation site. Nanocracks formed within the ordered grain boundary sample interior and grew until a bridging grain detached, connecting the internal nanocrack to the notch to allow the start of crack propagation. The AIF-containing sample had no grain growth at crack initiation and instead had distributed plastic activity surrounding the notch region. Similar to the ordered sample, internal cracking occurred until the bridging grain yielded, connecting the nanocrack to the notch and allowing crack propagation to commence.

2) *Propagation*: The ordered grain boundary sample demonstrated unsteady, accelerating crack growth characterized by the formation of an extensive nanocrack network interspersed with grain bridges. In contrast, the AIF-containing sample experienced steady, constant-rate crack growth with distributed plastic activity preceding the crack tip. The evenly distributed plastic activity in the AIF-containing sample indicates that the grain boundaries were better able to mitigate dislocation damage, whereas the ordered grain boundary sample had extensive nanocracking and highly localized plasticity.

In summary, the ordered grain boundary sample had highly localized plasticity with unsteady crack propagation and extensive nanocracking. The AIF sample instead demonstrated enhanced ductility with steady crack propagation and evenly distributed plastic activity, indicating that the AIFs diffused grain boundary strain and inhibited boundary fracture. While nanocrystalline metal grain sizes cause undesirable rapid crack growth during fatigue, the associated high volume fraction of grain boundaries may serve as a silver-lining. Extrinsically, grain bridging coupled with enhanced damage tolerance techniques can increase fatigue toughness by resisting crack propagation in nanocrystalline metals. Intrinsically, AIFs can diffuse grain boundary strain concentrations and promote dislocation activity for more stable crack growth. A simple thermal processing route has been shown to significantly enhance the ductile fatigue toughness of a nanocrystalline binary alloy suggesting a path forward for nanocrystalline alloys in fatigue related applications.

**Acknowledgements**

JDS and TJR were supported by the U.S. Department of Energy, Office of Basic Energy Sciences, Materials Science and Engineering Division under Award No. DE-SC0014232, and the U.S. Department of Energy, Office of Science, Office of Workforce Development for Teachers and Scientists, Office of Science Graduate Student Research (SCGSR) program. BLB, KH, CMB, and NMH were supported by the U.S. Department of Energy, Office of Basic Energy Sciences, Materials Science and Engineering Division under FWP 18-013170. The SCGSR program is administered by the Oak Ridge Institute for Science and Education for the DOE under contract number DE- SC0014664. TEM work was performed at the UC Irvine Materials Research Institute (IMRI). SEM and FIB work was performed at the UC Irvine Materials Research Institute (IMRI) using instrumentation funded in part by the National Science Foundation Center for Chemistry at the Space-Time Limit (CHE-0802913). Additional FIB and TEM work was performed at the Center for Integrated Nanotechnologies, an Office of Science User Facility operated for the U.S. Department of Energy (DOE) Office of Science. Sandia National Laboratories is a multi-mission laboratory managed and operated by National Technology and Engineering Solutions of Sandia, LLC., a wholly owned subsidiary of Honeywell International, Inc., for the U.S. DOE's National Nuclear Security Administration under contract DE-NA-0003525.The views expressed in the article do not necessarily represent the views of the U.S. Department of Energy or the United States Government.

## References

[1] K.M. Youssef, R.O. Scattergood, K.L. Murty, J.A. Horton, C.C. Koch, *Appl. Phys. Lett.*, 87, 091904 (2005).
[2] C.C. Koch, K.M. Youssef, R.O. Scattergood, K.L. Murty, *Adv. Eng. Mater.*, 7, 787 (2005).
[3] T.J. Rupert, C.A. Schuh, *Acta Mater.*, 58, 4137 (2010).
[4] R.O. Ritchie, *Int. J. Fracture*, 100, 55 (1999).
[5] T. Hanlon, E.D. Tabachnikova, S. Suresh, *Int. J. Fatigue*, 27, 1147 (2005).
[6] T. Hanlon, Y.N. Kwon, S. Suresh, *Scripta Mater.*, 49, 675 (2003).
[7] T.A. Furnish, D.C. Bufford, F. Ren, A. Mehta, K. Hattar, B.L. Boyce, *Scripta Mater.*, 143, 15 (2018).
[8] T.A. Furnish, A. Mehta, D. Van Campen, D.C. Bufford, K. Hattar, B.L. Boyce, *J. Mater. Sci.*, 52, 46 (2017).
[9] R.A. Meirom, D.H. Alsem, A.L. Romasco, T. Clark, R.G. Polcawich, J.S. Pulskamp, M. Dubey, R.O. Ritchie, C.L. Muhlstein, *Acta Mater.*, 59, 1141 (2011).
[10] M.D. Sangid, *Int. J. Fatigue*, 57, 58 (2013).
[11] H.A. Padilla, B.L. Boyce, *Exp. Mech.*, 50, 5 (2010).
[12] K.S. Kumar, H. Van Swygenhoven, S. Suresh, *Acta Mater.*, 51, 5743 (2003).
[13] F. Mompiou, M. Legros, A. Boé, M. Coulombier, J.P. Raskin, T. Pardoen, *Acta Mater.*, 61, 205 (2013).
[14] K.S. Kumar, S. Suresh, M.F. Chisholm, J.A. Horton, P. Wang, *Acta Mater.*, 51, 387 (2003).
[15] D. Gianola, B. Mendis, X. Cheng, K. Hemker, *Mater. Sci. Eng., A*, 483, 637 (2008).
[16] L. Wang, T. Xin, D. Kong, X. Shu, Y. Chen, H. Zhou, J. Teng, Z. Zhang, J. Zou, X.D. Han, *Scripta Mater.*, 134, 95 (2017).
[17] Z.X. Wu, Y.W. Zhang, M.H. Jhon, D.J. Srolovitz, *Acta Mater.*, 61, 5807 (2013).
[18] P. Liu, S.C. Mao, L.H. Wang, X.D. Han, Z. Zhang, *Scripta Mater.*, 64, 343 (2011).
[19] S.J. Dillon, M. Tang, W.C. Carter, M.P. Harmer, *Acta Mater.*, 55, 6208 (2007).
[20] G. Palumbo, S.J. Thorpe, K.T. Aust, *Scripta Met. et Mater.*, 24, 1347 (1990).
[21] A. Khalajhedayati, Z. Pan, T.J. Rupert, *Nat. Comm.*, 7, 10802 (2016).
[22] Z. Pan, T.J. Rupert, *Acta Mater.*, 89, 205 (2015).
[23] A. Khalajhedayati, T.J. Rupert, *JOM*, 67, 2788 (2015).
[24] L.F. Allard, W.C. Bigelow, M. Jose- Yacaman, D.P. Nackashi, J. Damiano, S.E. Mick, *Microsc. Res. Tech.*, 72, 208 (2009).
[25] L. Zhong, J. Wang, H. Sheng, Z. Zhang, S.X. Mao, *Nature*, 512, 177 (2014).
[26] D.C. Bufford, D. Stauffer, W.M. Mook, S.A. Syed Asif, B.L. Boyce, K. Hattar, *Nano Lett.*, 16, 4946 (2016).
[27] X.K. Zhu, *Intl. J. of Press. Vess. and Pip.*, 156, 40 (2017).
[28] V. Samaeeaghmiyoni, H. Idrissi, J. Groten, R. Schwaiger, D. Schryvers, *Micron*, 94, 66 (2017).
[29] K. Hattar, D.C. Bufford, D.L. Buller, *Nucl. Instrum. Methods Phys. Res., Sect. B*, 338, 56 (2014).
[30] S.K. Lin, Y.L. Lee, M.W. Lu, *Int. J. Fatigue*, 23, 75 (2001).
[31] R.C. Hugo, H. Kung, J.R. Weertman, R. Mitra, J.A. Knapp, D.M. Follstaedt, *Acta Mater.*, 51, 1937 (2003).
[32] C.M. Park, J.A. Bain, *J. Appl. Phys.*, 91, 6830 (2002).
[33] J.D. Schuler, T.J. Rupert, *Acta Mater.*, 140, 196 (2017).

[34] Q. Jin, D.S. Wilkinson, G.C. Weatherly, *J. Eur. Ceram. Soc.*, 18, 2281 (1998).
[35] J.D. Schuler, O.K. Donaldson, T.J. Rupert, *Scripta Mater.*, 154, 49 (2018).
[36] C.V. Thompson, *Annu. Rev. Mater. Sci.*, 20, 245 (1990).
[37] Y. Yang, B. Imasogie, G.J. Fan, P.K. Liaw, W.O. Soboyejo, *Metall. Mater. Trans. A*, 39, 1145 (2008).
[38] D. Farkas, M. Willemann, B. Hyde, *Phys. Rev. Lett.*, 94, 165502 (2005).
[39] R.O. Ritchie, *Mater. Sci. Eng., A*, 103, 15 (1988).
[40] J.K. Shang, R.O. Ritchie, *Metall. Trans. A*, 20, 897 (1989).
[41] R. Ramachandramoorthy, R. Bernal, H.D. Espinosa, *ACS Nano*, 9, 4675 (2015).
[42] Z. Liu, D. Yu, J. Tang, X. Chen, X. Wang, *Intl. J. of Press. Vess. and Pip.*, 168, 11 (2018).
[43] M.D. Thouless, *J. Am. Ceram. Soc.*, 71, 408 (1988).
[44] A. Vinogradov, *J. Mater. Sci.*, 42, 1797 (2007).
[45] R. Pippan, A. Hohenwarter, *Fatigue Fract. Eng. Mater. Struct.*, 40, 471 (2017).
[46] C. Laird, G. Smith, *Philosophical Magazine*, 7, 847 (1962).
[47] J. Xie, X. Wu, Y. Hong, *Scripta Mater.*, 57, 5 (2007).
[48] K. Tanaka, Y. Nakai, M. Yamashita, *Int. J. Fracture*, 17, 519 (1981).
[49] C. Bjerkén, S. Melin, *Eng. Frac. Mech.*, 71, 2215 (2004).
[50] M.Y. Gutkin, I. Ovid'ko, *Philos. Mag. Lett.*, 84, 655 (2004).
[51] S. Suresh, *Metall. Trans. A*, 14, 2375 (1983).
[52] A.M. Gokhale, W.J. Drury, S. Mishra, Recent Developments in Quantitative Fractography, *Fractography of Modern Engineering Materials: Composites and Metals*, ed. J.E. Masters, L.N. Gilbertson, (Philadelphia: ASTM, 1993), p. 3.
[53] D. Farkas, H. Van Swygenhoven, P. Derlet, *Phys. Rev. B: Condens. Matter*, 66, 060101 (2002).
[54] M.D. Sangid, G.J. Pataky, H. Sehitoglu, R.G. Rateick, T. Niendorf, H.J. Maier, *Acta Mater.*, 59, 7340 (2011).
[55] Y. Zhang, G.J. Tucker, J.R. Trelewicz, *Acta Mater.*, 131, 39 (2017).
[56] R. Liu, Y. Tian, Z. Zhang, X. An, P. Zhang, Z. Zhang, *Scientific Reports*, 6, 27433 (2016).
[57] G.Q. Xu, M.J. Demkowicz, *Phys. Rev. Lett.*, 111, 145501 (2013).
[58] T. Leitner, A. Hohenwarter, R. Pippan, *Mater. Sci. Eng., A*, 646, 294 (2015).
[59] T.R. Bieler, P. Eisenlohr, F. Roters, D. Kumar, D.E. Mason, M.A. Crimp, D. Raabe, *Int. J. Plast.*, 25, 1655 (2009).
[60] A. Singh, L. Tang, M. Dao, L. Lu, S. Suresh, *Acta Mater.*, 59, 2437 (2011).
[61] L. Liu, J. Wang, S.K. Gong, S.X. Mao, *Scientific Reports*, 4, 4397 (2014).
[62] V. Turlo, T.J. Rupert, *Acta Mater.*, 151, 100 (2018).

| | **Avg. Deflection Length (nm)** | | **Avg. Deflection Angle (°)** | | **Tortuosity** | |
|---|---|---|---|---|---|---|
| | **Propagation** | **Failure** | **Propagation** | **Failure** | **Propagation** | **Failure** |
| **Ordered Grain Boundary** | 37.7 | 34.3 | 68.8 | 83.3 | 1.53 | 1.53 |
| **AIF-Containing** | 46.2 | 38.9 | 58.9 | 82.2 | 1.01 | 1.52 |

**Table I**. Fracture surface analysis of the ordered grain boundary and amorphous intergranular film (AIF) containing samples from the propagation and failure stages of the fatigue crack lifetime.

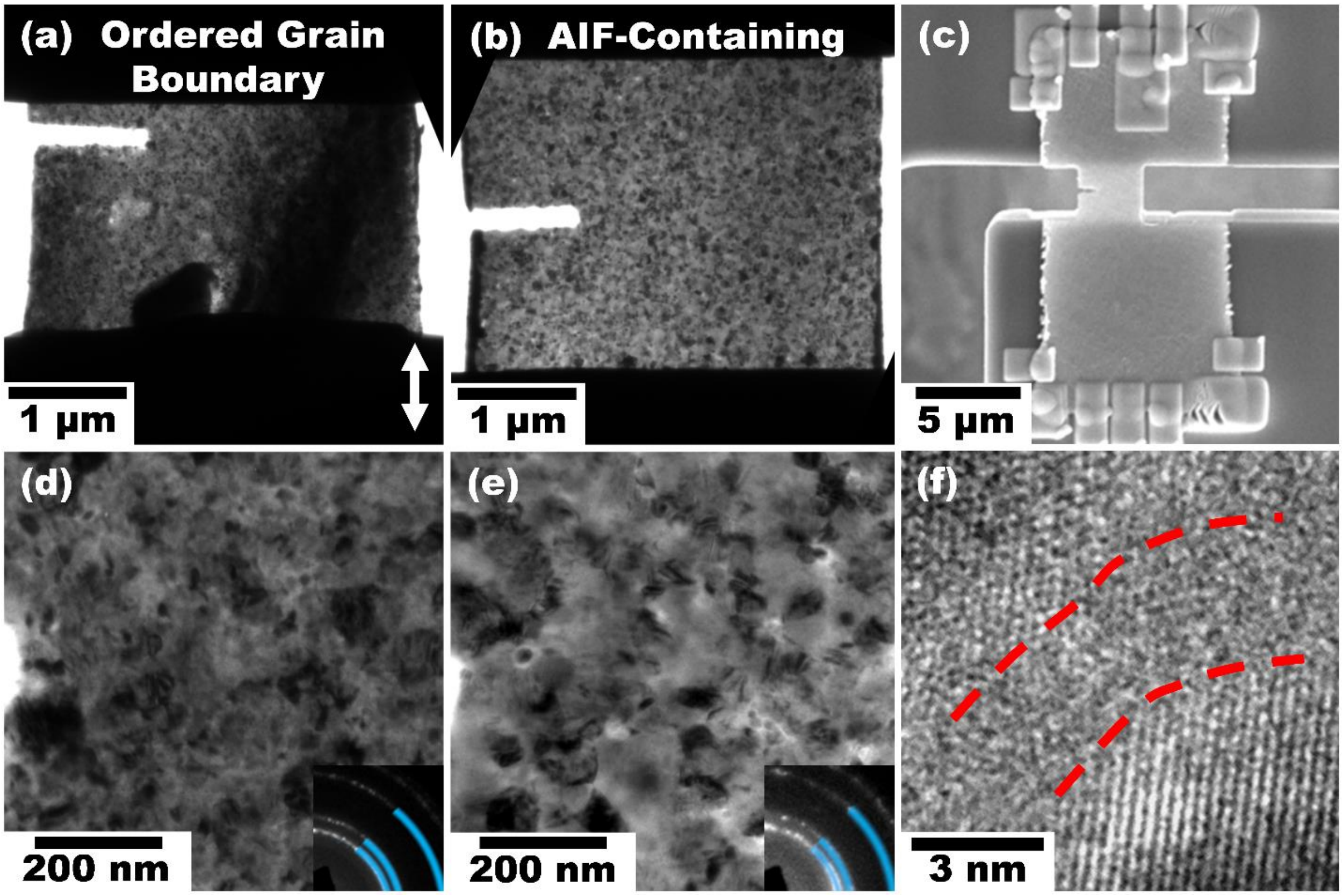

**Fig. 1**. Bright field transmission electron microscopy (TEM) images of the Cu-1 at.% Zr films with (a) ordered grain boundaries and (b) amorphous intergranular films (AIFs). (c) Scanning electron microscopy image of the fatigue sample preparation on the push-to-pull device. Bright field TEM images of the region adjacent to the notch (left) are shown for the (d) ordered grain boundary and (e) AIF-containing samples. The insets in (d) and (e) show the corresponding electron diffraction patterns with the Cu face-centered cubic rings superimposed in blue. (f) High resolution TEM image showing an AIF from the AIF-containing sample outlined by dashed red lines.

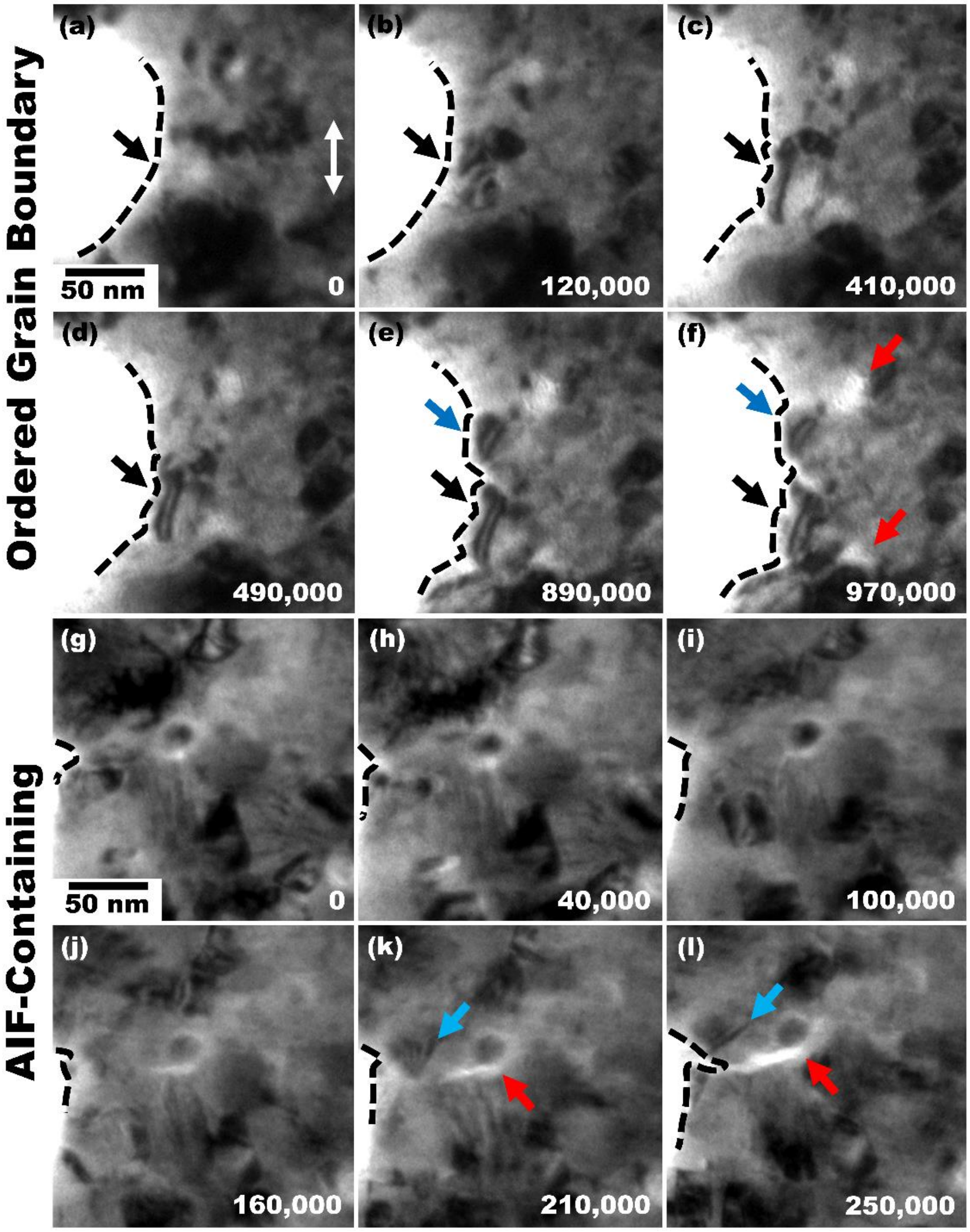

**Fig. 2**. Bright field transmisison electron microscopy images of the (a)-(f) ordered grain boundary and (g)-(l) amorphous intergranular film (AIF) containing samples that show crack initiation with progressive cycling, where the number of elasped cycles is indicated in the bottom right. The dashed lines show the notch. The black arrows in (a)-(f) identify a grain that grows with cycling and where crack initiation eventually occurs, while the blue arrows show a competing crack initiation site. The red arrows in (f) indicate intergranular cracks formed in front of the notch at the competing crack initiation sites. The blue arrows in (k) and (l) indicate a grain that plastically deforms and yields due to the nucleating crack, and the red arrows show an interior crack formed in front of the notch. The white arrow in (a) shows the loading direction.

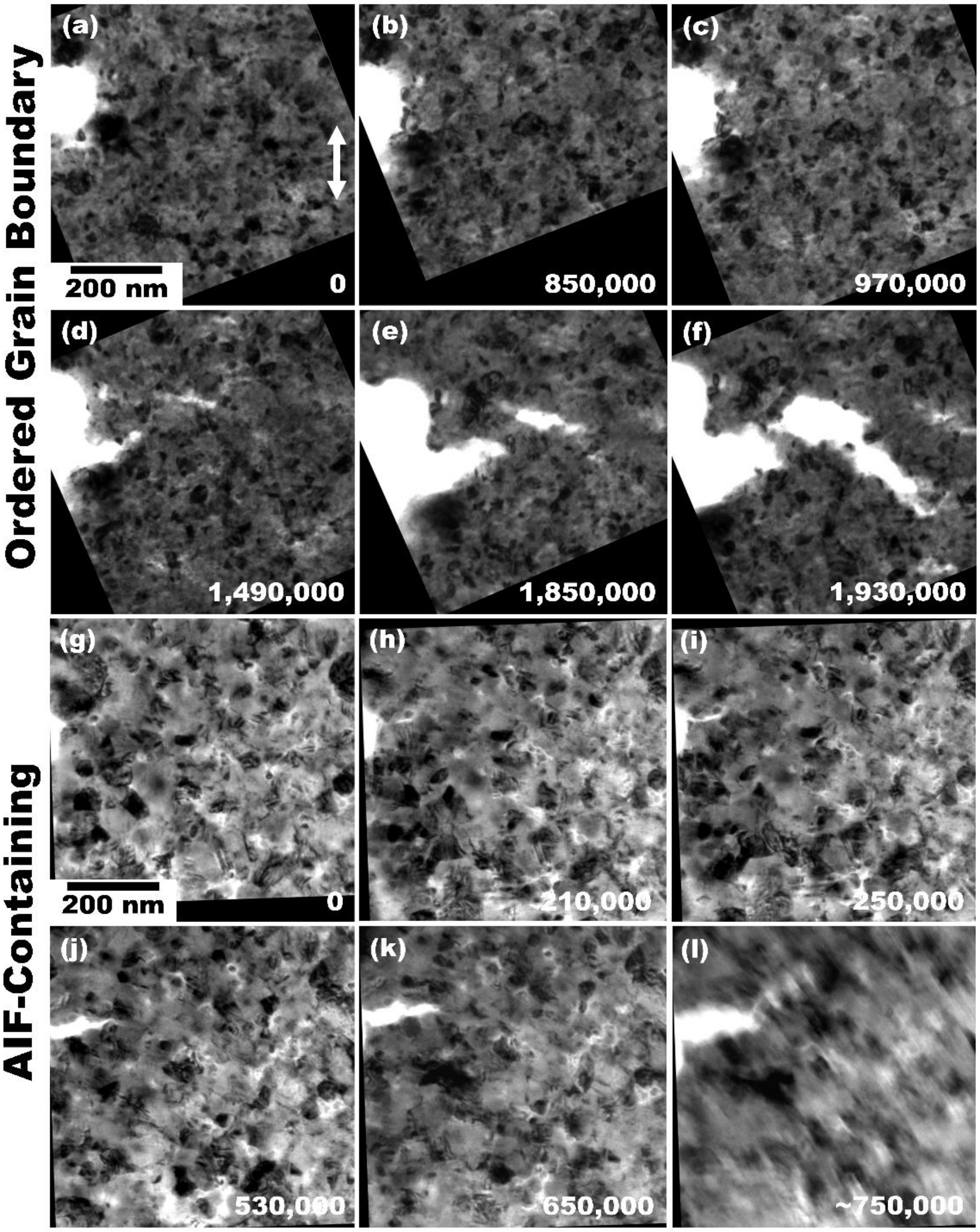

**Fig. 3**. Bright field transmission electron microscopy images of the (a)-(f) ordered grain boundary and (g)-(l) amorphous intergranular film (AIF) containing samples showing crack propagation with progressive fatigue cycling, where the number of elapsed cycles is indicated in the bottom right. The white arrow in (a) shows the loading direction. The last frame before failure is shown in (f) and (l), where the (f) ordered grain boundary sample failed during constant loading between fatigue cycling and the (l) AIF-containing sample failed during fatigue cycling, causing the image to be blurry.

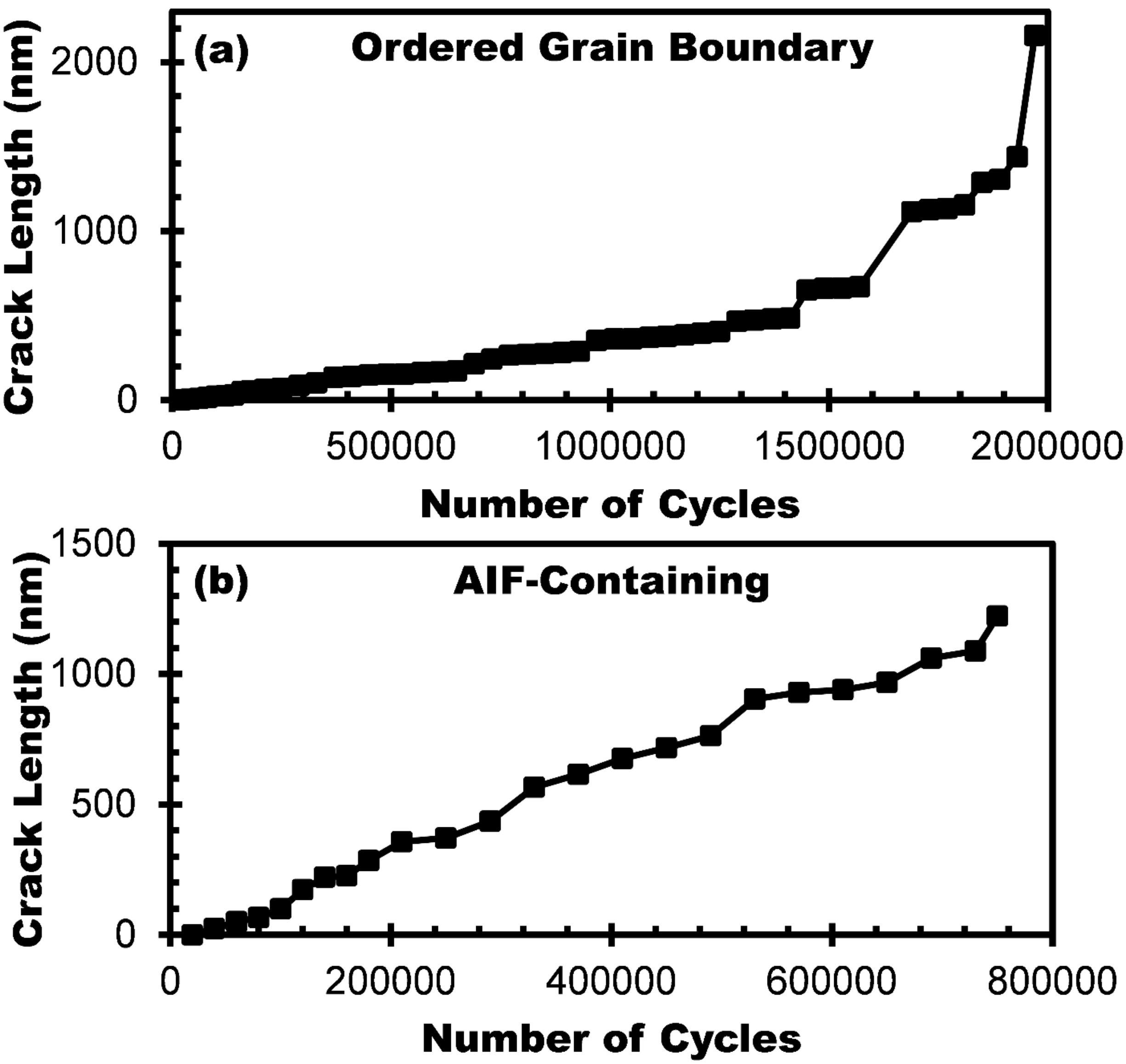

**Fig. 4**. Crack growth beginning at the notch as a function of fatigue cycle for the (a) ordered grain boundary and (b) amorphous intergranular film (AIF) containing samples.

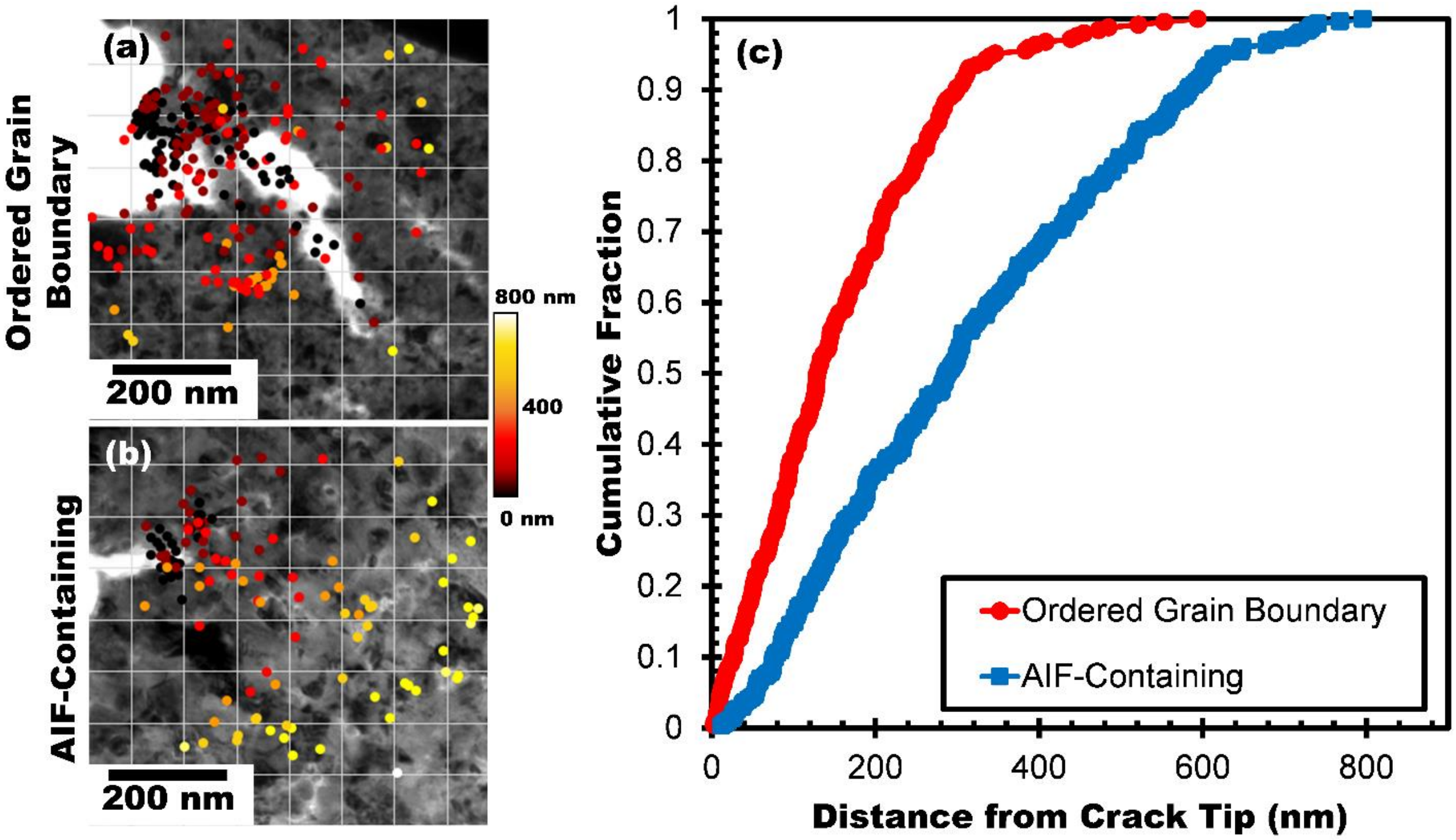

**Fig. 5**. Heat maps of the (a) ordered grain boundary and (b) amorphous intergranular film (AIF) containing samples showing the location of plastic activity identified dynamically in frame by frame analysis through the video in front of the advancing crack tip accumulated throughout the fatigue tests. The color gradient shows the distance of the plastic event from the crack tip at the time of detection, where black is closest and white is farthest. The backgrounds are bright field transmission electron microscopy images of the last clear frame before fracture for reference. (c) The cumulative distribution fraction of plastic activity as a function of the distance from the crack tip at the point of detection. The ordered grain boundary sample data is shown with red circles and the AIF-containing sample data appears as blue squares.

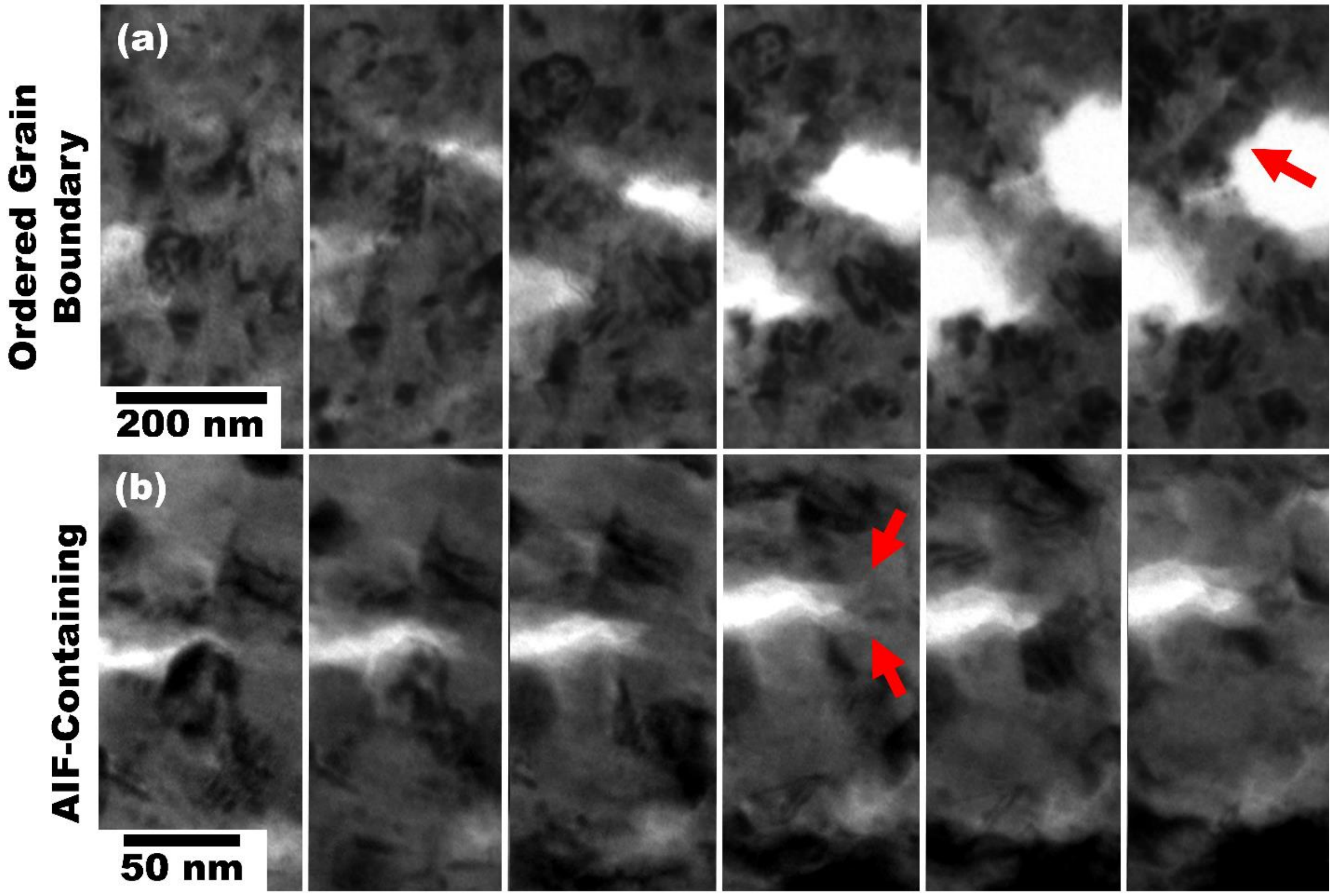

**Fig. 6**. Bright field transmission electron microscopy images of the (a) ordered grain boundary and (b) amorphous intergranular film (AIF) containing samples showing the evolution of microstructrual events that occurred in each film during fatigue. A grain bridging event in the ordered grain boundary sample that sustained substantial plastic activity is visible in (a), where the grain indicated by the red arrow grew across the bridge and served as the eventual point of detachment. The plastic deformation zone in front of the crack tip of the AIF-containing sample appears in (b), where a distinctive "V" shape preceding the crack is highlighted by the red arrows.

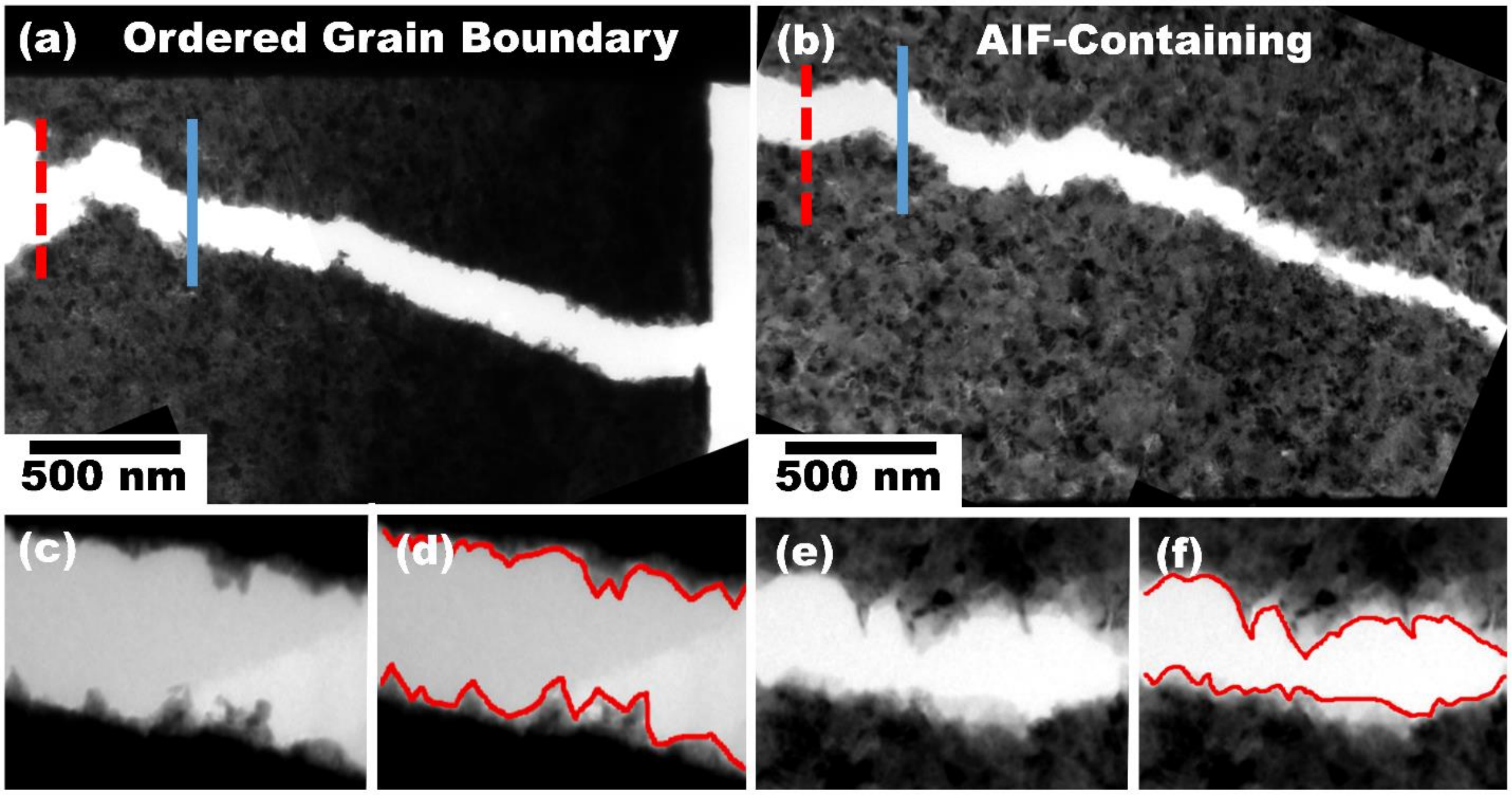

**Fig. 7**. Bright field transmission electron microscopy images are shown of the (a) ordered grain boundary and (b) amorphous intergranular film (AIF) containing samples after fatigue failure. The dashed red lines indicate the point of crack initiation where crack propagation occurs to the right until reaching the solid blue line, where sudden failure then commenced. The fracture surfaces are shown in greater detail for the (c) ordered grain boundary and (e) AIF-containing samples, with outlines of the fracture surface shown for each sample in (d) and (f).

# Supplementary Information

## Supplementary Note 1

Cu-1 at.% Zr thin films were sputter deposited onto NaCl polished disc substrates 13 mm × 1 mm in dimension (International Crystal Laboratories, Inc. #0002A-4549). Portions of the film were then floated by dissolution of the substrate in a solution of water and isopropyl alcohol onto Protochips, Inc. Fusion heating chips with no supporting membrane (Protochips, Inc. #E-FHBN-10), where special care taken to not short the opposing electrical leads on either side of the heating chip window with the floated film. A small amount of Cu oxide was seen in some samples after high vacuum annealing to create the ordered and disordered samples and may be due to film interactions with the salt substrate or dissolution in water.

After annealing, portions of the film located on the heating chips were prepared for in situ TEM fatigue using a focused ion beam (FIB) lift-out technique. Portions of the annealed films chosen for lift-out were in the center of the heating chip with good contact to the window to ensure adequate thermal transfer. Supplementary Fig. 1(a) shows a scanning electron microscopy (SEM) image of the thin film on a heating chip over the window holes with good thermal contact. Supplementary Fig. 1(b) shows rectangles approximately 15 nm × 15 nm cut using FIB and lifted out through Pt attachment to an Omniprobe, Inc. micro-manipulator visible on the left. Next, the samples were placed across the 2.5 μm wide gauge section of the push-to-pull (PTP) device indicated by the red arrow in Supplementary Fig. 1(c) and secured with electron and ion beam Pt, with final FIB sample and notch shaping shown in Supplementary Fig. 1(d). Special care was taken during Pt deposition to (1) maximize the distance of the attaching Pt from the gauge section and (2) avoid imaging after deposition until base pressure was recovered in order to minimize inadvertent Pt deposition over the gauge section. If these measures were not taken, enough accidental Pt deposition occurred to destroy electron transparency over the gauge area and make mechanical testing of the experimental specimen inconclusive. The PTP devices are from the Bruker Corporation with a stiffness of 450 N/m (Bruker Corporation #5-1092-HIGH-10) and were attached to the Bruker PI 95 PicoIndenter TEM holder Cu mount using conductive silver paint. Gluing was completed before lift-out of the sample from the heating chip onto the PTP device in order to minimize handling of the completed sample, where even the slightest vibration or pressure on the PTP device can break the extremely delicate sample once stretched across the gauge section.

The methodology used minimizes FIB damage since the deposited film thickness was sufficiently electron transparent with no further FIB thinning required, indicated by the increased brightness of the thin film over the gauge region in Supplementary Fig. 1(d). Also, the thin film is in the correct orientation for transfer from the heating chip to the PTP device, which eliminates the geometric challenge faced when using a standard vertical lift-out technique for TEM sample preparation to create a plan view specimen. Other methods to minimize FIB damage or achieve a plan view lift-out have been investigated in the literature [1-3].

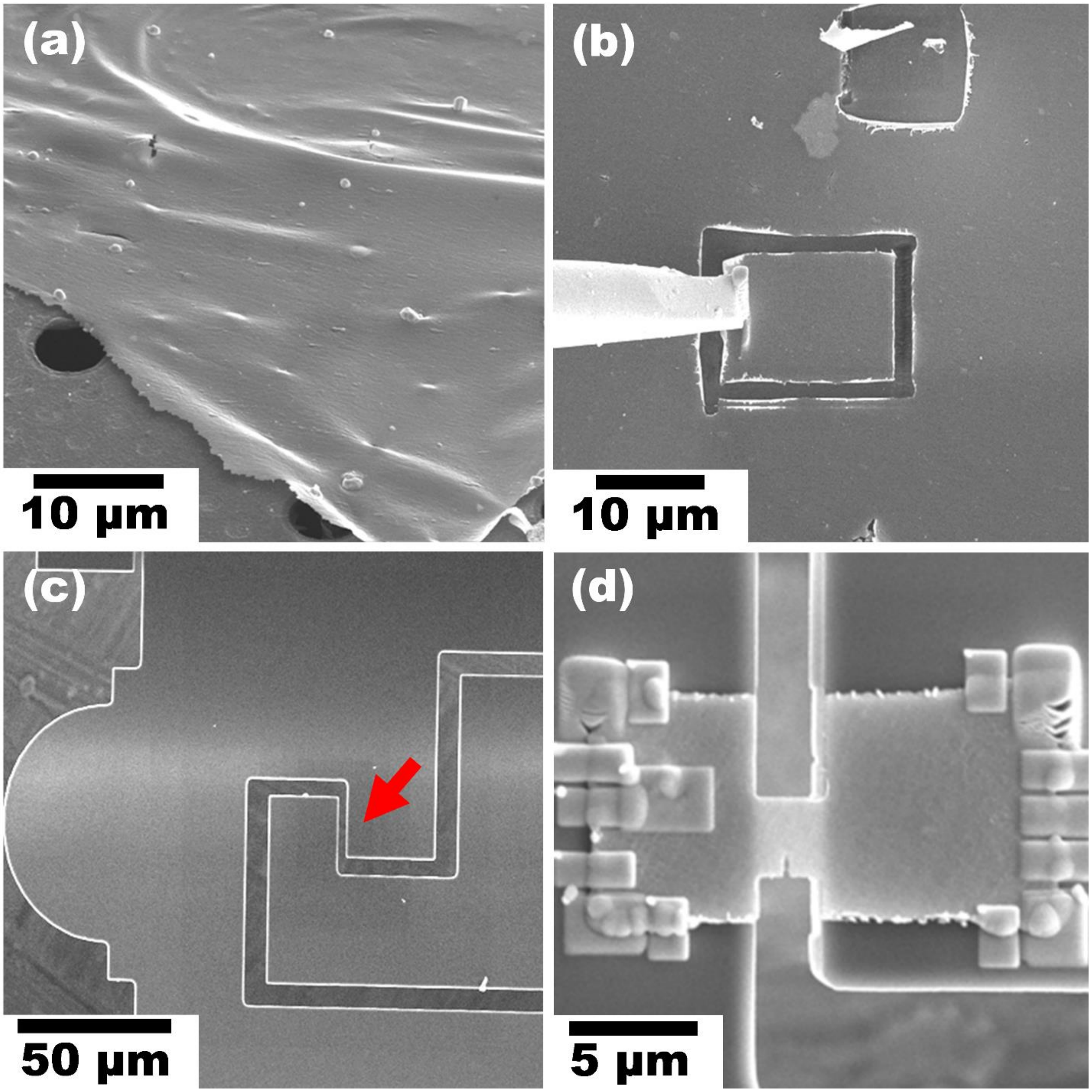

**Supplementary Fig. 1.** (a) A scanning electron microscopy image of the Cu-1 at.% Zr thin film that has been floated onto the heating chip window area where the heating chip holes are visible. (b) A selected portion of the annealed thin film cut into a rectangle using the focused ion beam (FIB) technique and Pt attached to the micro-manipulator on the left for lift-out. (c) The push-to-pull device with the gauge section indicated by the red arrow. (d) The finished specimen, attached with Pt and shaped using the FIB.

## Supplementary Note 2

Electron energy loss spectroscopy (EELS) was performed using a JEOL GrandARM300CF in scanning transmission electron microscopy (STEM) mode operated at 300 kV in order to compute the thickness of the as-deposited Cu-1 at.% Zr thin film. The thickness was calculated using the log-ratio (absolute) method with a measured convergence semi-angle of 31.0 mrad and collection semi-angle of 39.4 mrad. An effective atomic number of 29 was used due to the low Zr dopant concentration. The as-deposited thin film is shown using bright field transmission electron microscopy (TEM) in Supplementary Fig. 2(a) and high angle annular dark field STEM in Supplementary Fig. 2(b), where the dashed red line in Supplementary Fig. 2(b) shows where the thickness measurement was performed. Supplementary Fig. 2(c) shows the thickness profile with an average film thickness of 51 ± 6 nm.

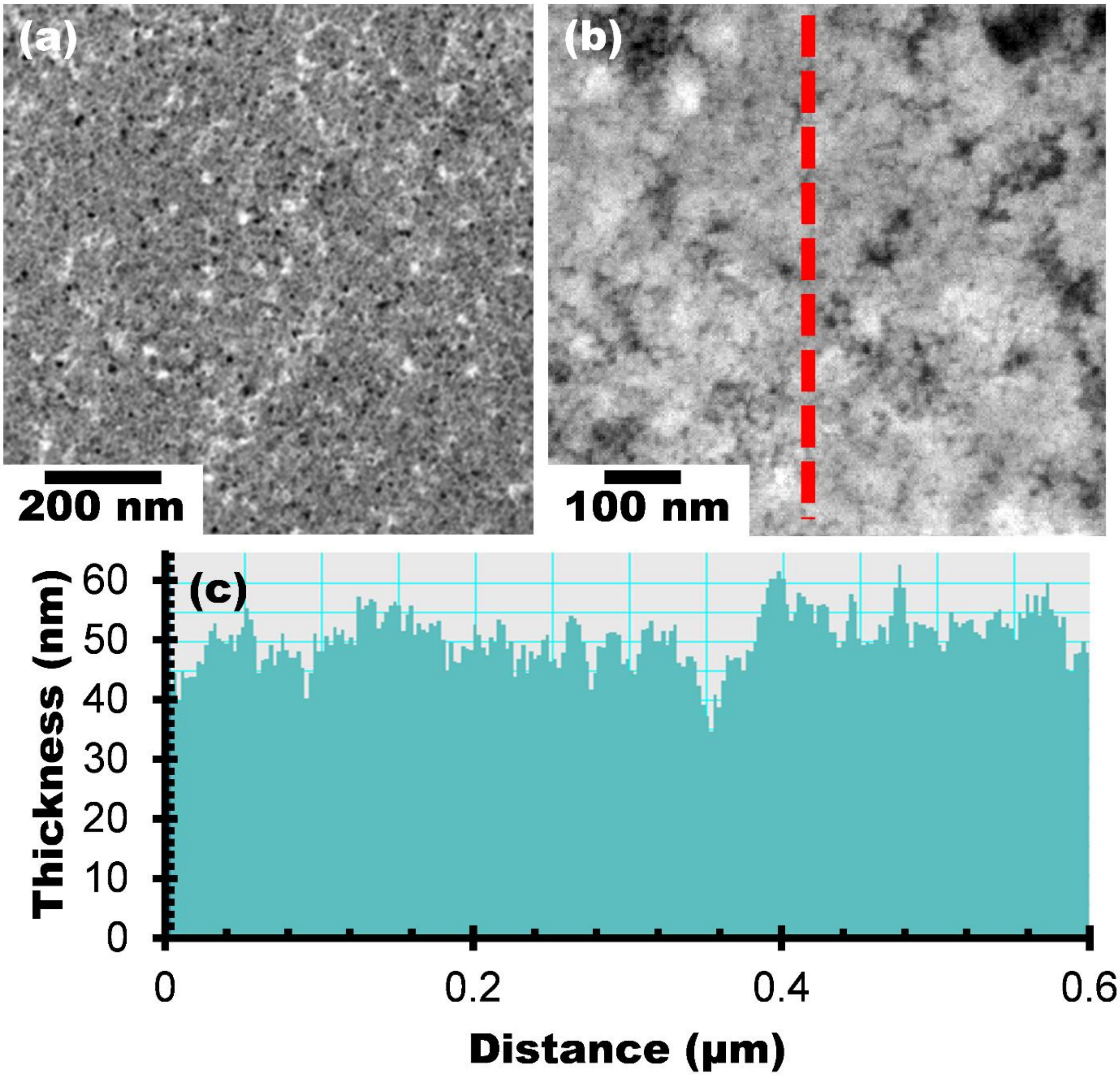

**Supplementary Fig. 2.** The as-deposited Cu-1 at.% Zr thin film is shown using (a) bright field transmission electron microscopy and (b) high angle annular dark field scanning transmission electron microscopy. The film thickness was measured at the dashed red line in (b), with data presented in (c).

**Supplementary Note 3**

The full in situ TEM fatigue loading conditions and crack growth data for the ordered grain boundary and AIF-containing samples are presented in Supplementary Tables I and II, where sample fracture occurred at the final steps. Reported loads are on the total system including both the PTP device and specimen. The notch plus crack length, *a*, is normalized by the specimen width, *W*. *da/dN* is the crack growth rate measured as the change in crack length per number of cycles. Step 45 of the ordered sample was performed at a lower total mean load than step 44 due to the fatigue test being paused and then resumed. A representation of these loading conditions is shown in Supplementary Fig. 3 for the ordered grain boundary sample where the total mean load is plotted and the vertical bars represent the total load amplitude. Supplementary Fig. 4 shows the loading conditions within an individual step in greater detail using step 17 from the AIF-containing sample as an example for the load and displacement versus step time duration in Supplementary Figs. 4(a) and (b) respectively. The 200 Hz loading rate causes the curves to appear as solid black boxes.

| **Ordered Grain Boundary Sample** | | | | | | | |
|---|---|---|---|---|---|---|---|
| **Step** | **Frequency (Hz)** | **Duration (s)** | **Mean Load, Total (μN)** | **Load Amplitude, Total (μN)** | **Crack Length, *a* (μm)** | **Normalized Crack Length, *a/W*** | **Crack Growth Rate, *da/dN* (m/cyc)** |
| 0 | 100 | 200 | 10 | 5 | 1.000 | 0.303 | NA |
| 1 | 100 | 200 | 10 | 5 | 1.004 | 0.304 | NA |
| 2 | 100 | 200 | 20 | 5 | 1.009 | 0.306 | 2.8E-13 |
| 3 | 100 | 200 | 30 | 10 | 1.015 | 0.308 | 2.7E-13 |
| 4 | 100 | 200 | 40 | 10 | 1.022 | 0.310 | 2.4E-13 |
| 5 | 100 | 200 | 50 | 15 | 1.024 | 0.310 | 3.9E-13 |
| 6 | 100 | 200 | 60 | 20 | 1.028 | 0.312 | 4.6E-13 |
| 7 | 100 | 200 | 70 | 20 | 1.051 | 0.318 | 4.6E-13 |
| 8 | 100 | 200 | 80 | 30 | 1.055 | 0.320 | 3.2E-13 |
| 9 | 150 | 200 | 80 | 30 | 1.062 | 0.322 | 2.6E-13 |
| 10 | 200 | 200 | 80 | 30 | 1.069 | 0.324 | 3.1E-13 |
| 11 | 200 | 200 | 80 | 30 | 1.086 | 0.329 | 4.4E-13 |
| 12 | 200 | 200 | 80 | 30 | 1.102 | 0.334 | 4.7E-13 |
| 13 | 200 | 200 | 80 | 30 | 1.134 | 0.344 | 4.0E-13 |
| 14 | 200 | 200 | 90 | 30 | 1.139 | 0.345 | 2.8E-13 |
| 15 | 200 | 200 | 90 | 30 | 1.148 | 0.348 | 1.2E-13 |
| 16 | 200 | 200 | 90 | 30 | 1.151 | 0.349 | 1.3E-13 |
| 17 | 200 | 200 | 100 | 30 | 1.153 | 0.349 | 1.2E-13 |
| 18 | 200 | 200 | 100 | 30 | 1.162 | 0.352 | 1.4E-13 |
| 19 | 200 | 200 | 100 | 30 | 1.165 | 0.353 | 3.3E-13 |
| 20 | 200 | 200 | 110 | 30 | 1.172 | 0.355 | 5.2E-13 |
| 21 | 200 | 200 | 110 | 30 | 1.214 | 0.368 | 6.7E-13 |
| 22 | 200 | 200 | 110 | 30 | 1.242 | 0.376 | 6.1E-13 |
| 23 | 200 | 200 | 120 | 30 | 1.265 | 0.383 | 3.7E-13 |
| 24 | 200 | 200 | 120 | 30 | 1.270 | 0.385 | 2.1E-13 |
| 25 | 200 | 200 | 120 | 30 | 1.273 | 0.386 | 1.4E-13 |
| 26 | 200 | 200 | 130 | 30 | 1.280 | 0.388 | 4.5E-13 |
| 27 | 200 | 200 | 130 | 30 | 1.288 | 0.390 | 6.2E-13 |
| 28 | 200 | 200 | 130 | 30 | 1.352 | 0.410 | 5.9E-13 |
| 29 | 200 | 200 | 140 | 30 | 1.360 | 0.412 | 4.3E-13 |
| 30 | 200 | 200 | 140 | 30 | 1.362 | 0.413 | 1.4E-13 |
| 31 | 200 | 200 | 140 | 30 | 1.370 | 0.415 | 1.5E-13 |
| 32 | 200 | 200 | 150 | 30 | 1.376 | 0.417 | 2.0E-13 |
| 33 | 200 | 200 | 150 | 30 | 1.384 | 0.419 | 2.2E-13 |
| 34 | 200 | 200 | 150 | 30 | 1.396 | 0.423 | 5.0E-13 |
| 35 | 200 | 200 | 160 | 30 | 1.405 | 0.426 | 6.2E-13 |
| 36 | 200 | 200 | 160 | 30 | 1.466 | 0.444 | 5.9E-13 |
| 37 | 200 | 200 | 160 | 30 | 1.472 | 0.446 | 4.3E-13 |
| 38 | 200 | 200 | 170 | 30 | 1.479 | 0.448 | 9.5E-13 |
| 39 | 200 | 200 | 170 | 30 | 1.485 | 0.450 | 1.4E-12 |
| 40 | 200 | 200 | 170 | 30 | 1.650 | 0.500 | 1.4E-12 |
| 41 | 200 | 200 | 180 | 30 | 1.660 | 0.503 | 9.6E-13 |
| 42 | 200 | 200 | 180 | 30 | 1.662 | 0.504 | 2.0E-12 |
| 43 | 200 | 200 | 180 | 30 | 1.670 | 0.506 | 2.3E-12 |
| 44 | 200 | 200 | 180 | 30 | NA | NA | NA |
| 45 | 200 | 200 | 80 | 30 | NA | NA | NA |
| 46 | 200 | 200 | 90 | 30 | 2.115 | 0.641 | 2.3E-12 |
| 47 | 200 | 200 | 100 | 30 | 2.126 | 0.644 | 2.0E-12 |
| 48 | 200 | 200 | 110 | 30 | 2.132 | 0.646 | 9.3E-13 |
| 49 | 200 | 200 | 120 | 30 | 2.155 | 0.653 | 1.3E-12 |
| 50 | 200 | 200 | 130 | 30 | 2.287 | 0.693 | 1.9E-12 |
| 51 | 200 | 200 | 140 | 30 | 2.306 | 0.699 | 5.4E-12 |
| 52 | 200 | 200 | 150 | 30 | 2.440 | 0.739 | NA |
| 53 | 200 | 200 | 160 | 30 | 3.158 | 0.957 | NA |

**Supplementary Table I.** The loading parameters per step of the ordered grain boundary sample in situ transmission electron microscopy fatigue and associated crack growth data.

| AIF-Containing Sample | | | | | | | |
|---|---|---|---|---|---|---|---|
| **Step** | **Frequency (Hz)** | **Duration (s)** | **Mean Load, Total (μN)** | **Load Amplitude, Total (μN)** | **Crack Length, *a* (μm)** | **Normalized Crack Length, *a/W*** | **Crack Growth Rate, *da/dN* (m/cyc)** |
| 0 | 100 | 200 | 10 | 5 | 1.000 | 0.294 | NA |
| 1 | 100 | 200 | 10 | 5 | 1.024 | 0.301 | NA |
| 2 | 100 | 200 | 20 | 5 | 1.051 | 0.309 | 1.2E-12 |
| 3 | 100 | 200 | 30 | 10 | 1.066 | 0.314 | 1.7E-12 |
| 4 | 100 | 200 | 40 | 10 | 1.100 | 0.323 | 2.2E-12 |
| 5 | 100 | 200 | 50 | 15 | 1.173 | 0.345 | 2.2E-12 |
| 6 | 100 | 200 | 60 | 20 | 1.221 | 0.359 | 2.1E-12 |
| 7 | 100 | 200 | 70 | 20 | 1.226 | 0.361 | 2.0E-12 |
| 8 | 100 | 200 | 80 | 30 | 1.285 | 0.378 | 1.6E-12 |
| 9 | 150 | 200 | 80 | 30 | 1.357 | 0.399 | 1.5E-12 |
| 10 | 200 | 200 | 80 | 30 | 1.372 | 0.403 | 1.7E-12 |
| 11 | 200 | 200 | 80 | 30 | 1.436 | 0.422 | 1.8E-12 |
| 12 | 200 | 200 | 80 | 30 | 1.566 | 0.461 | 2.0E-12 |
| 13 | 200 | 200 | 80 | 30 | 1.616 | 0.475 | 1.7E-12 |
| 14 | 200 | 200 | 90 | 30 | 1.676 | 0.493 | 1.2E-12 |
| 15 | 200 | 200 | 90 | 30 | 1.717 | 0.505 | 1.7E-12 |
| 16 | 200 | 200 | 90 | 30 | 1.763 | 0.519 | 1.7E-12 |
| 17 | 200 | 200 | 100 | 30 | 1.905 | 0.560 | 1.5E-12 |
| 18 | 200 | 200 | 100 | 30 | 1.929 | 0.567 | 1.1E-12 |
| 19 | 200 | 200 | 100 | 30 | 1.940 | 0.571 | 8.8E-13 |
| 20 | 200 | 200 | 110 | 30 | 1.968 | 0.579 | 1.1E-12 |
| 21 | 200 | 200 | 110 | 30 | 2.062 | 0.606 | 1.8E-12 |
| 22 | 200 | 200 | 110 | 30 | 2.088 | 0.614 | NA |
| 23 | 200 | 200 | 120 | 30 | 2.223 | 0.654 | NA |

**Supplementary Table II.** The loading parameters per step of the amorphous intergranular film (AIF) containing sample in situ transmission electron microscopy fatigue and associated crack growth data.

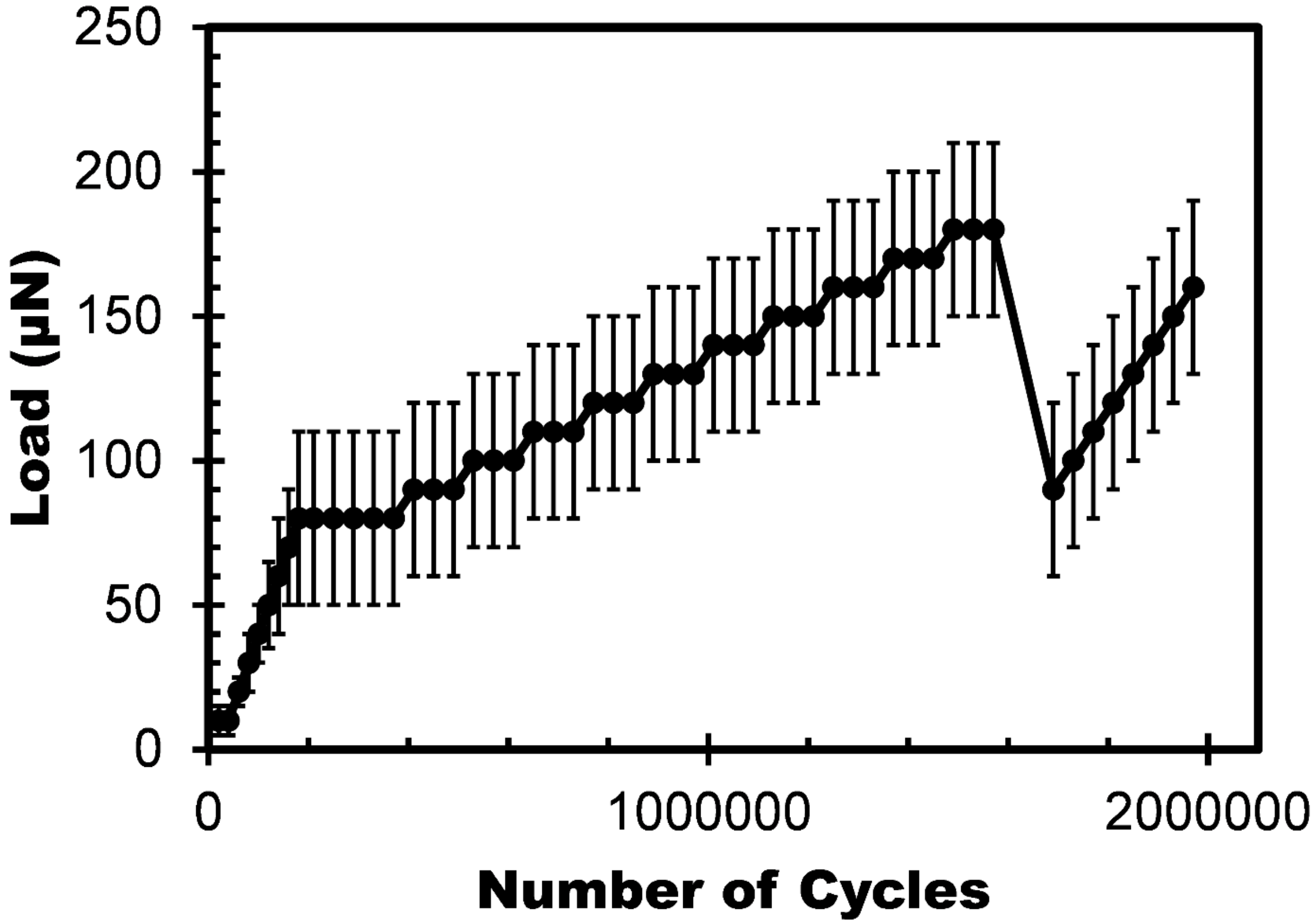

**Supplementary Fig. 3.** The loading conditions of the ordered grain boundary sample for each step are shown where the total mean load is plotted and the vertical bars represent the total load amplitude for each step. The drop at 1,690,000 is due to the fatigue cycling being paused and then resumed. The amorphous intergranular film containing sample received identical loading conditions through to its point of failure.

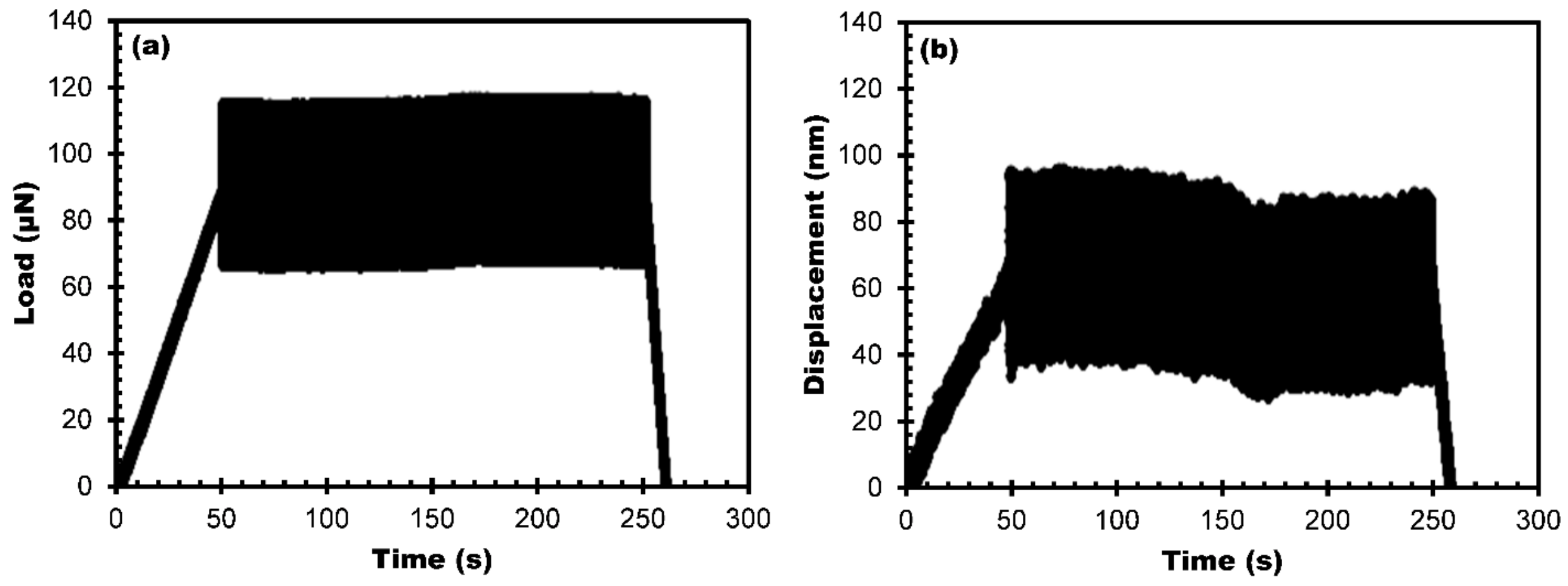

**Supplementary Fig. 4.** Representative (a) load and (b) displacement curves from the amorphous intergranular film fatigue cycling at loading step number 17. The 200 Hz loading rate causes the curves to appear as solid black boxes.

Supplementary Video 1 shows the in situ TEM fatigue for the ordered grain boundary sample.

Supplementary Video 2 shows the in situ TEM fatigue for the amorphous intergranular film (AIF) containing sample.

## Supplementary References

[1] J.P. Liebig, M. Göken, G. Richter, M. Mačković, T. Przybilla, E. Spiecker, O.N. Pierron, B. Merle, *Ultramicroscopy*, 171, 82 (2016).
[2] C. Li, G. Habler, L.C. Baldwin, R. Abart, *Ultramicroscopy*, 184, 310 (2018).
[3] V. Samaeeaghmiyoni, H. Idrissi, J. Groten, R. Schwaiger, D. Schryvers, *Micron*, 94, 66 (2017).